\documentclass[twocolumn]{aastex631} 

\usepackage{graphics,epsf}
\usepackage[utf8]{inputenc}
\usepackage{amsmath}                
\usepackage{amsfonts}               
\usepackage{amssymb}                
\usepackage{epsfig}                 
\usepackage{graphicx}
\usepackage{float}
\usepackage{color}
\usepackage{multirow}               
\usepackage{hyperref}
\usepackage{xspace}

\hypersetup{
    colorlinks=true,
    linkcolor=red,   
    urlcolor=cyan}

\usepackage[para,online,flushleft]{threeparttable}

\usepackage[colorinlistoftodos]{todonotes}

\newcommand{\mesa}{\textsc{mesa}}

\newcommand{\cm}{{~\rm cm}}
\newcommand{\km}{{~\rm km}}
\newcommand{\s}{{~\rm s}}

\newcommand{\g}{{~\rm g}}
\newcommand{\G}{{~\rm G}}
\newcommand{\K}{{~\rm K}}
\newcommand{\erg}{{~\rm erg}}

\newcommand{\Mo}{~M_\odot}

\begin{document}

\title{The jittering jets explosion mechanism (JJEM) in electron capture supernovae}
\date{June 2023}
\author[0009-0004-9646-5271]{Nikki Yat Ning Wang}
\affiliation{Department of Physics, Technion, Haifa, 3200003, Israel; nikki.wang.19@ucl.ac.uk; s.dmitry@campus.technion.ac.il; soker@physics.technion.ac.il}

\author[0000-0002-9444-9460]{Dmitry Shishkin}
\affiliation{Department of Physics, Technion, Haifa, 3200003, Israel; nikki.wang.19@ucl.ac.uk; s.dmitry@campus.technion.ac.il; soker@physics.technion.ac.il}

\author[0000-0003-0375-8987]{Noam Soker}
\affiliation{Department of Physics, Technion, Haifa, 3200003, Israel; nikki.wang.19@ucl.ac.uk; s.dmitry@campus.technion.ac.il; soker@physics.technion.ac.il}

\begin{abstract}
We conduct one-dimensional stellar-evolution simulations of stars with zero-age main sequence masses of $M_{\rm ZAMS}= 8.8-9.45 M_\odot$ towards core collapse by electron capture and find that the convective zone of the pre-collapse core can supply the required stochastic angular momentum fluctuations to set a jet-driven electron capture supernova (ECSN) explosion in the frame of the jittering jets explosion mechanism (JJEM). By our assumed criteria of minimum convective specific angular momentum and an accreted mass during jet-launching of $M_{\rm acc} \simeq 0.001-0.01 M_\odot$, the layer in the convective zone that when accreted launches the exploding jittering jets resides in the helium-rich zone. Depending on the model, this exploding layer is accreted at about a minute to a few hours after core collapse occurs, much shorter than the time the exploding shock crosses the star.  The final (gravitational) mass of the neutron star (NS) remnant is in the range of $M_{\rm NS} =1.25-1.43 M_\odot$.
\end{abstract} 

\section{INTRODUCTION}
\label{sec:Intro}

Stars with zero-age main sequence (ZAMS) mass of $M_{\rm{ZAMS}}\gtrsim 10\Mo$, depending on metalicity, develop an iron (iron group elements) core of $M_{\rm Fe} \simeq 1.2 M_\odot$ and undergo (iron) core-collapse supernova (CCSN) with typical explosion energies of $E_{\rm ex} \approx 10^{51} \erg$ and leave either a neutron star (NS) or a black hole (BH) remnant. These are termed FeCCSNe.   

Stars in the intermediate mass range $\Delta M_{\rm ECSN}$ between those that leave a white dwarf remnant and those that undergo FeCCSN, $8 \Mo \lesssim M_{\rm{ZAMS}} \lesssim 10\Mo$ where the limits depend on the initial metallicity (e.g., \citealt{Doherty_etal_2015, Doherty_etal_2017,Chanlaridis_etal_2022, Cinquegrana_etal_2023}), are presumed to collapse as an electron capture supernova (ECSNe, e.g., \citealt{Miyaji_etal_1980}). The actual mass range for ECSNe is smaller than the above range, i.e., $\Delta M_{\rm ECSN} < 1M_\odot$. ECSNe do not develop a massive iron core but rather explode when the main isotopes of their core are $^{16}$O, $^{20}$Ne, $^{28}$Si, $^{32}$S, and some $^{24}$Mg; iron group elements account for less than half of the core mass.
The electron capture processes on $^{24}$Mg and then $^{20}$Ne disturb the hydrostatic equilibrium in the inner core and the core collapses to form a neutron star in an ECSNe (e.g., \citealt{Miyaji_etal_1980, Nomoto_1984, Nomoto_1987, Poelarends_etal_2008, Leung_Nomoto_2019, Zha_etal_2019, GuoYLetal2024, Limongi_etal_2024}). The typical explosion energy of ECSNe is expected to be $E_{\rm ex} \approx 10^{49} - 10^{50} \erg$ and they leave an NS remnant. An example is probably the Crab Nebula with an explosion energy of $E_{\rm ex} \simeq 10^{50} \erg$   (see analysis by \citealt{Yang_Chevalier_2015}). 

We note that there are no unequivocal observational identifications of ECSNe. However, there are some candidates. One promising candidate is SN 2018zd \citep{Hiramatsuetal2021}, with an explosion energy of $\simeq 2 \times 10^{50} \erg$. \cite{Hiramatsuetal2021} crudely estimate that 0.6-8.5\% of all CCSNe are ECSNe, corresponding to a progenitor mass range of only $\Delta M_{\rm ECSN} \simeq 0.06-0.69M_\odot$.

In CCSNe (both FeCCSNe and ECSNe), the explosion energy comes from a small fraction of the gravitational energy that the collapsing core releases.  The collapsing core forms a central proto-NS at nuclear densities that stops the collapse, i.e., the core bounce phase, and drives a shock wave that propagates out. In FeCCSNe the shock stalled at $r \simeq 150 \km$. There are two alternative mechanisms to channel a small fraction of the gravitational energy to explode the star, the delayed neutrino explosion mechanism and the jittering jets explosion mechanism (JJEM), as we summarize in section \ref{sec:Two Mechanism}.

Motivated by recent conclusions (e.g., \citealt{BearSoker2024, Soker2024CFpoint}) that many CCSN remnants are shaped by jets with varying directions, most likely related to the jets that exploded the star in the JJEM (section \ref{sec:Two Mechanism}; for a review see \citealt{Soker2024Rev}), we examine whether the pre-collapse convective layers in the outer core and above the core of progenitors of ECSNe can supply sufficiently large angular momentum fluctuations as the JJEM requires. In section \ref{sec:NumScheme}, we describe our numerical setting and assumptions, and in sections \ref{sec:M88} - \ref{sec:Models}, we describe the results for several stellar models. We summarize this study in section \ref{sec:Summary}  

\section{The two explosion mechanisms}
\label{sec:Two Mechanism}

According to the delayed neutrino explosion mechanism, heating by neutrinos over a second to several seconds revives the shock; the re-expanding shock explodes the star  (\citealt{BetheWilson1985}, with hundreds of papers in the four decades since then, e.g., \citealt{Hegeretal2003, Janka2012, Nordhausetal2012, Gabayetal2015, Ertl_etal_2016, Mulleretal2019Jittering, Bruenn_etal_2020_neutrinoExplosionCode, Bollig_etal_2021, Burrows_etal_2020_simulations, Stockingeretal2020, Fujibayashietal2021, Fryeretal2022,  Nakamuraetal2022, Olejaketal2022, Bocciolietal2023, Burrowsetal2023, BoccioliRoberti2024}).
The neutrino-driven explosion mechanism encounters several challenges. One is that research groups disagree on both qualitative and quantitative results.\footnote{In a review talk in the \textit{Transients Down Under} meeting, held in Melbourne, Australia on 29 January 2024, Hans-Thomas Janka writes in the summary of his talk about the delayed neutrino explosion mechanism:  ``3D models of different groups disagree: there are qualitative and quantitative differences!''.} For instance, \cite{Radice_etal_2017} find an explosion time of $t \simeq 0.6 \s$ after the collapse with a leftover proto-neutron star mass of $\rm{M_{NS}} \simeq 1.2 \Mo$ and explosion energy to range from $1.2-1.8 \times 10^{50} \erg$ for the models that do explode. For the same ECSN progenitor model of \cite{Nomoto_1984,Nomoto_1987}, simulations by other groups have resulted in explosion energies that are substantially different from \cite{Radice_etal_2017}, e.g., \cite{Wanajo_etal_2003} find explosion energies of $0.2-35 \times 10^{50} \erg$,  \cite{Kitaura_etal_2006} find $\approx 10^{50} \erg$, and \cite{Janka_etal_2008} $< 10^{50} \erg$. Moreover, not all of the models provided by the neutrino-driven manage to explode. For instance, out of the 24 models in \cite{Radice_etal_2017}, only 15 of them exploded, with the mass range from $\rm{\Mo}=9.0, 10.0, 11.0$ not exploding in the 1D simulations. Another challenge of the neutrino-driven explosion mechanism is to explain the point-symmetric morphology that was identified in the last year in several CCSNe remnants, and which suggests multi-pair jet activity at explosion (e.g., \citealt{BearSoker2024, Soker2024CFpoint}).

In the JJEM \citep{Soker2010}, on the other hand, jets that the newly born neutron star launches in stochastically varying directions, i.e., jittering jets,  explode the star (e.g., \citealt{Papish2011_jittering, PapishSoker2014, GilkisSoker2014, GilkisSoker2015, GilkisSoker2016, Quataertetal2019, Soker2019_SASIjets, Soker2020RAA, Soker2023gap, Soker20231987A, ShishkinSoker2021, ShishkinSoker2022, ShishkinSoker2023, AntoniQuataert2022, AntoniQuataert2023}). 
Neutrino heating does play a role in the JJEM by boosting the jet-driven explosion \citep{Soker2022_Boosting}. 
The accreted gas with stochastically varying angular momentum, in both magnitude and direction, forms the intermittent accretion disks that launch the jittering jets from the newly born NS. The initial source of the stochastic angular momentum is the stochastic motion of convective layers in the pre-collapse core that introduces seed angular momentum perturbations. The angular momentum fluctuations due to the random velocities of the convective cells are amplified by instabilities behind the stalled shock at $r \lesssim 100 \km$ above the newly born NS. The spiral standing accretion shock instability (spiral SASI) is the dominant instability that we expect to amplify the seed angular momentum perturbations of the accreted convective layers (for the SASI see, e.g., \citealt{BlondinMezzacappa2007, Iwakamietal2014, Kurodaetal2014, Fernandez2015, Kazeronietal2017}; for spiral-SASI see, e.g., \citealt{Andresenetal2019, Walketal2020, Nagakuraetal2021, Shibagakietal2021}). Other types of instabilities are possible. However, to fully explore the instabilities and processes that operate in the JJEM, the simulations must include magnetic fields (e.g., \citealt{Soker2018arXivB, Soker2019RAAB, Soker2020RAAB}). 

Backed-up by the new finding of multi-pair jet activity in CCSN remnants (e.g., \citealt{Soker2024Rev, Soker2024CFpoint}), we here examine whether the criteria that were used for the operation of the JJEM in FeCCSNe also apply to ECSNe.

To form an accretion disk around an NS, the specific angular momentum of the accreted matter $j_{\rm acc}$ should be about equal or larger than the specific angular momentum of a particle in a centrifugally-supported accretion disk on the surface of the NS $j_{\rm d,NS}$, i.e., $j_{\rm acc} \gtrsim j_{\rm d,NS} \simeq 2 \times 10^{16} \cm^2 \s^{-1}$ (e.g. \citealt{PapishGilkisSoker20158}). Several convective cells contribute the material to the accreted gas at any time. At a large fraction of the accretion process, according to the JJEM, the specific angular momentum after amplification by the spiral-SASI is sufficiently large to form a short-lived ($\simeq 0.01-0.1 \s$) accretion disk around the NS. We define the angular momentum parameter as 
\begin{equation}
    j(r) \equiv r v_{\rm conv}, 
\label{eq:j}
\end{equation} 
where $r$ is the radius of the layer in the pre-collapse core and $v_{\rm conv}$ is the convective velocity as the one-dimensional stellar evolutionary code gives according to the mixing length theory (MLT). Therefore, the specific angular momentum of the gas formed by several cells will be smaller than $j$, the value of one convective cell. 
If there is sufficient amplification of the specific angular momentum fluctuations by the spiral-SASI or another type of instability, then, at sporadic periods, the specific angular momentum of the accreted gas is sufficiently large to form an accretion disk (or an accretion belt) around the NS. Therefore, for launching jittering jets, the JJEM conjectures that the value of $j$ can be smaller than $j_{\rm d,NS}$. This conjecture has not been shown yet in independent simulations, but 3D simulations of the delayed-neutrino mechanism have hinted at a factor $\simeq 3$ scaling between one-dimensional MLT convection and that in hydrodynamical simulations (e.g., \citealt{FieldsCouch2020}), further softening this assumption. We follow \cite{ShishkinSoker2022} who study FeCCSNe, and assume the constraint for ECSNe to be the same, i.e., $j \gtrsim 0.1-0.25 j_{\rm d,NS} \simeq 2 \times 10^{15} - 5 \times 10^{15} \cm^2 \s^{-1}$. Earlier studies, in particular  \cite{ShishkinSoker2022}, concluded that progenitors of FeCCSNe have sufficiently large angular momentum fluctuations, i.e., $j > 2 \times 10^{15} \cm^2 \s^{-1}$, in their convective zones near silicon burning and/or oxygen burning to support the JJEM.  

To summarize, we emphasize that the criterion of $j \gtrsim 0.1-0.25 j_{\rm d,NS} \simeq 2 \times 10^{15} - 5 \times 10^{15} \cm^2 \s^{-1}$ is at present an assumption of the JJEM. We here follow this earlier assumption that has been applied only to FeCCSNe (e.g., \citealt{ShishkinSoker2021} based on \citealt{PapishGilkisSoker20158}), and apply it to ECSNe.  It will have to be confirmed by future studies. On the other hand, the JJEM has received very strong support from studies in the last year that identified point-symmetric morphologies in several CCSNe (e.g., \citealt{Soker2024Rev, Soker2024CFpoint}), in particular in the iconic CCSN remnant Cassiopeia A (\citealt{BearSoker2024}). The comparison of CCSN remnants with jet-inflated structures in clusters and groups of galaxies strongly suggests that pairs of jets also shaped many CCSN remnants, and in several of these by two and more pairs of opposite jets (\citealt{Soker2024CFpoint}). As stated above, these types of morphologies are expected in the JJEM and pose severe challenges to the neutrino-explosion mechanism. 
Nonetheless, the formation of accretions disks from fluctuations that add up to a total zero angular momentum is the main theoretical challenge of the JJEM. While previously quantified \citep{GilkisSoker2014, GilkisSoker2015}, this reduction in the stochastic angular momentum of the material reaching the spiral SASI modes remains a hindering factor. Confronting this challenge requires future studies to use magneto-hydrodynamical simulations constructed specifically to search for the formation of intermittent accretion disks from turbulent convective matter.

\section{Numerical Scheme}
\label{sec:NumScheme}

The simulations of the ECSN progenitors were conducted with the one-dimensional stellar evolution code \textsc{mesa} (Modules for Experiments in Stellar Astrophysics; version r23.05.1; \citealt{Paxton2011, Paxton2013, Paxton2015, Paxton2018, Paxton2019, Jermyn2023}). Several non-rotating models of ECSN progenitors are calculated with references to the initial mass and metallicity in \cite{Doherty_etal_2015} and \cite{Poelarends_etal_2008}. Two models will be presented in detail and referred to as M88 and M91. Model M88 starts with an initial (zero-age main sequence; ZAMS), mass of $M_{\rm ZAMS} = 8.8M_\odot$ and metallicity of $z = 0.004$, whereas model M91 starts with $M_{\rm ZAMS} = 9.1M_\odot$ at $z = 0.0086$. The outcomes of other models will be summarized in a  table. 

We chose the masses and metalicities of the five models that we simulate because these are presumed to end as ECSNe (e.g., \citealt{Poelarends_etal_2008, Jones_etal_2013, Doherty_etal_2015}), specifically in the range between $M_{\rm ZAMS} = 9.45M_\odot, z = 0.014$ and $M_{\rm ZAMS} = 8.8M_\odot, z=0.004$. Based on these numbers, we extrapolated a linear mass-metallicity relationship that supposedly produces ECSNe progenitors between these coordinates. The linear relationship between mass and metallicity of ECSNe can be inferred from \citealt{Doherty_etal_2015}.

We use the 'Dutch' wind loss scheme (e.g., \citealt{Vink2001,NugisLamers2000,Glebbeek2009}) with a scaling factor of 1, the Skye equations-of-state (\citealt{Jermyn2021}), and the Type 2 OPAL opacity table (\citealt{Iglesias1996}).

The outcomes of the simulations of the final phases of the ECSN progenitors, just before collapse, sensitively depend on numerical parameters (e.g., \citealt{Chanlaridis_etal_2022,Woosley_Heger_2015,Doherty_etal_2015}). Altering some parameters, such as the convection and overshooting scheme can result in different stellar remnants. To present an ECSN model, we use a similar treatment to that by \cite{Chanlaridis_etal_2022} for convection and coulomb correction. This includes using the Ledoux criterion (\citealt{Henyey_etal_1965}) with mixing length coefficient $\alpha_{\rm MLT}=2$, semi-convection coefficient $\alpha_{\rm SC} = 1$, and thermohaline coefficient $\alpha_{\rm thermo}= 1$. The MLT option here is selected to be TDC (time-dependent convection), which converges to the MLT described in \cite{Cox&Giuli_1968} in the long term \citep{Jermyn2023}. The change in energy caused by changing ion charges and changing electron chemical potential, namely \textsc{ion\_coulomb\_corrections} and \textsc{electron\_coulomb\_corrections}, are accounted for by numerical parameters depicted in \cite{Potekhin_etal_2009} and \cite{Itoh_etal_2002}. During the later stages of evolution for all models, the star experiences thermal pulses. Therefore, a strict overshooting scheme is needed to avoid numerical difficulties (e.g., \citealt{Jones_etal_2013}). We chose to use an exponential overshooting scheme, as described in \cite{Herwig2000}, and the overshoot parameters to be $f=0.009$ and $f_{0}=0.005$ in general; $f=0.007$ and $f_{0}=0.005$ for convective region near metal burning shells - as recommended in \cite{Jones_etal_2013}. Thus overshooting is $0.004 \cdot H_{p}$ and $0.002\cdot H_{p}$, with $H_{p}$ being the pressure scale height, over the convective boundary of convection zones in general and metal burning convection zones, respectively. 

We use a custom large nuclear network of 132 isotopes up to $^{60}$Zn (see Fig.~\ref{fig:nucNet} in Appendix) to account for weak reactions, namely, electron capture and Urca processes, that arise after the formation of the ONeMg core (e.g., \citealt{Jones_etal_2013, Takahashi_etal_2013}). The weak rates for the atomic mass of A=20 to A=28 isotopes tabulated by \cite{Suzuki_2016} are used to account for reactions relevant to high-density ONeMg cores.

The models go through an evolution in the super asymptotic giant branch phase, during which they experience thermal pulses and strong stellar winds that cause numerical difficulties in the envelope. 
Therefore, in the later stages of evolution, we set the tau\_factor parameter to the higher value of $\rm tau\_factor = 15$ (than the default photosphere edge value of $\rm tau\_factor = 1$). We do this to negate the influence of the numerical difficulties on the progression of the simulation. While this may alter the properties of the outer envelope we are mostly interested in the region near the core.

Inlists and a selected profile of the M88 (Section \ref{sec:M88}) model are available online, at \dataset[doi:10.5281/zenodo.10827141]{https://doi.org/10.5281/zenodo.10827141}.

\section{The $M_{\rm ZAMS} = 8.8M_\odot$ model (M88)}
\label{sec:M88}
\subsection{General properties}
\label{subsec:GeneralProperties88}

We examine the properties of the convection layer in the outer parts of the core and above the core as close as we can get the simulation to electron-capture core collapse. We do not study the entire chain of events that lead to the collapse of the core which can be found in other studies (e.g., \citealt{Jones_etal_2013, Woosley_Heger_2015, Takahashi_etal_2013, Nomoto_1987, Miyaji_etal_1980, Leung_Nomoto_2019,Zha_etal_2019, Limongi_etal_2024}). These, and other studies, find that in the late evolutionary stages of an ECSN progenitor close to collapse, processes driven by the weak nuclear force, such as the Urca processes and electron capture (mainly onto $^{24}$Mg), remove energy from the center of the core. As a result, the core contracts, and oxygen and neon start to burn off-center, producing heavier elements, mainly $^{28}$Si and $^{32}$S.
The O+Ne off-center burning can be described as a laminar flame propagating inwards (for treatments of laminar flames see, e.g., \citealt{Timmes2000, Leung_Nomoto_2019, Jones_etal_2013, Jones_etal_2016}). As the deflagration flame approaches the center, it triggers further electron capture reactions (mainly onto $^{20}$Ne) and induces collapse (\citealt{Takahashi_etal_2013,Woosley_Heger_2015}). 
The burning can reach iron peak elements (e.g., \citealt{Leung_Nomoto_2019, Takahashi_etal_2013}),
but ECSNe explode before they develop a massive iron core. Note that during this process, the contraction of the core is in constant competition with oxygen or sometimes neon deflagration, which could lead to a thermonuclear runaway (e.g., \citealt{Miyaji_etal_1980, Nomoto_1987, Leung_Nomoto_2019}).

There have been ongoing discussions about whether the final fate of a star within the ECSNe mass range will collapse into a neutron star or undergo thermonuclear runaway (e.g. \citealt{Miyaji_etal_1980, Jones_etal_2016, Zha_etal_2019}). \cite{Jones_etal_2016} show that some candidates for ECSNe with insufficient semiconvection and low ignition density for oxygen deflagration will undergo thermonuclear runaway instead. However, \cite{Zha_etal_2019} shows, by using the most updated electron capture rate by \cite{Suzuki_etal_2019}, that since the ONeMg core has oxygen deflagration at a high density ($\log \rho_{\rm core} > 10.01$) anyways, stars with a degenerate ONeMg core are likely to collapse as ECSNe.

All our models, as ECSNe progenitors from the super asymptotic giant branch channel, are expected to experience $10^4-10^5$ years of thermal pulsing until they reach electron capture or lose all of their envelopes (\citealt{Doherty_etal_2017}). 
We simulate the first few thermal pulses for each model until the numerical time steps become too low and the simulations practically stop. 
\cite{Limongi_etal_2024} advanced further into the thermal pulsing phase and found a positive linear relationship between the CO core mass and time during this phase. This indicates that the process behind the carbon-oxygen (CO) core growth, being helium shell burning, is somewhat constant. Hence we assume that both the CO core mass and the properties of the relevant convection zones, which are above the core, change rather insignificantly until the core collapses. 
A more detailed description of the thermal pulsing phase is given in \cite{Doherty_etal_2015,Doherty_etal_2017}.
The relevant point for our goals here (as we discuss later) is that the density of the envelope above the core where the relevant convective zones are, is very low and their exact location does not change the final NS mass.  

\begin{figure*}[t]
\begin{center}
\includegraphics[trim=0.5cm 0cm 0.5cm 1.65cm,width=\textwidth,scale=.66]{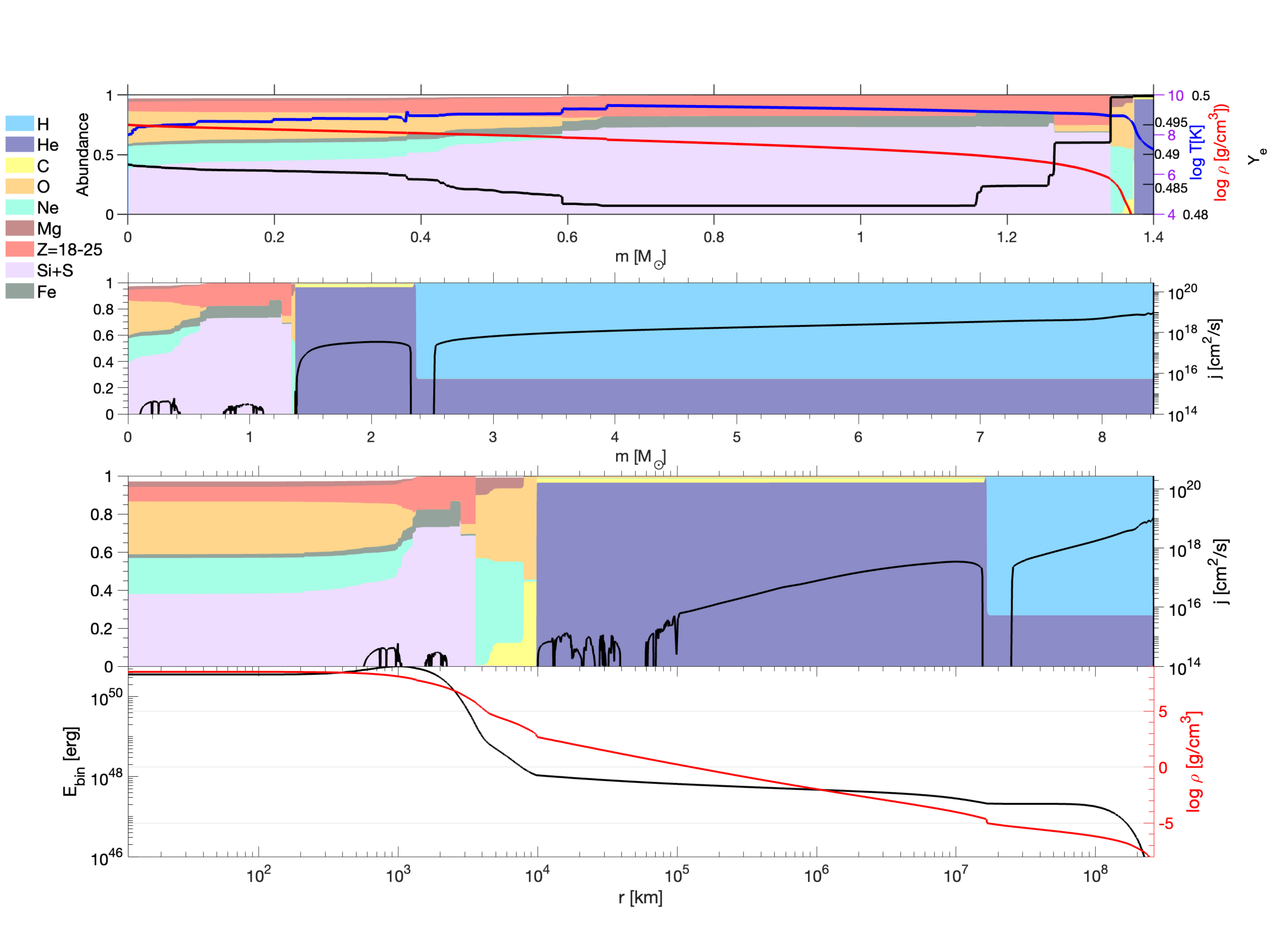} 
\caption{Some properties of the M88 model, i.e., $M_{\rm ZAMS} = 8.8 M_\odot$, at the last available time of our simulation ($t=0$) as a function of the mass and of the radius. Upper panel: The composition according to the legend on the left, the logarithms of the temperature (blue) and density (red), and the electron fraction $Y_{\rm e}$ (black), all as a function of mass coordinate $m$ in the inner $1.4 M_\odot$, roughly around the boundary of the CO core. Second panel: The composition as a function of the mass coordinate of the entire star, with a profile of the angular momentum parameter $j(r)$ (equation \ref{eq:j}). Lower two panels: The composition, $j(r)$, the binding energy $E_{\rm bin}(r)$ of mass residing above $r$, and the density (log scale) as functions of the radial coordinate in the star. }
\label{fig:FinalCompM88}
\end{center}
\end{figure*}

In Figure \ref{fig:FinalCompM88} we present several properties of model M88 at the last simulated time. 
In the upper two panels, we present the composition, temperature, density, and electron fraction $Y_{\rm e}$ (defined as the number of free electrons per baryon), in the inner $1.4 M_\odot$, and in the second panel we present the composition and angular momentum parameter $j(r)$ (equation \ref{eq:j}) as functions of mass in the entire star. In the third and fourth panels, we present the composition, $j(r)$, density, and binding energy above radius $r$, $E_{\rm bin}(r)$, as functions of the radial coordinate inside the star.

A prominent composition feature is a substantial off-center fraction of silicon and sulfur (Si+S).  The fraction of the iron group is meager. Another prominent composition feature is the large fraction of neon in the center. Electron capture starts on magnesium and then on neon. Overall, the composition in the core and the density at the center of the core of $\rho_{\rm core} \simeq 3 \times 10^8 \g \cm^{-3}$ are compatible with the early stage of pre-core-collapse due to electron capture.
The composition change from a region being O-Ne rich to it being Si+S-rich over time, for instance at $m \simeq 0.6-1.3 M_\odot$, leads to a lowering of $\rm Y_e$ at that region. The nucleosynthesis of heavier elements in the center facilitates the evolution towards electron capture by inducing contraction.

Due to numerical difficulties, we do not reach actual core collapse. Based on the core composition at the last simulation time, and comparison to \cite{Doherty_etal_2015}, \cite{Jones_etal_2013}, \cite{Stockingeretal2020}, and \cite{Limongi_etal_2024} that we referred to above, we infer that the relevant convective zones (see below) above the core in our last stellar model are not likely to change much until the full collapse. There is still some evolution of the inner core to collapse, like towards nuclear statistical equilibrium (NSE) in the inner core. The $8.8\Mo$ model with a solar composition that \cite{Stockingeretal2020} present just at core collapse has an inner mass of $0.45 \Mo$ in NSE, and a CO core radius of only $\simeq 1200 \km$. Also, they do not have an extended helium-rich layer as we have here, of about $1\Mo$. However, the exact structure depends on the helium core mass. \cite{Hillebrandtetal1984} find that in their simulation starting with a helium core mass of $2.2\Mo$, at a time of 100 years before the collapse, there is a very thin helium-rich layer, while when they start with a helium core mass of $2.4\Mo$ there is a thick helium layer, $\simeq 1\Mo$ at that time. Our helium core mass is $\simeq 2.4\Mo$, and therefore we expect to have a thick helium layer, as we do get here.

\subsection{The relevant convective zones}
\label{subsec:ConvectiveZones88}

\begin{figure*}[t]
\begin{center}
\includegraphics[trim=0.5cm 0cm 0.5cm 0.57cm,width=\textwidth,scale=.6]{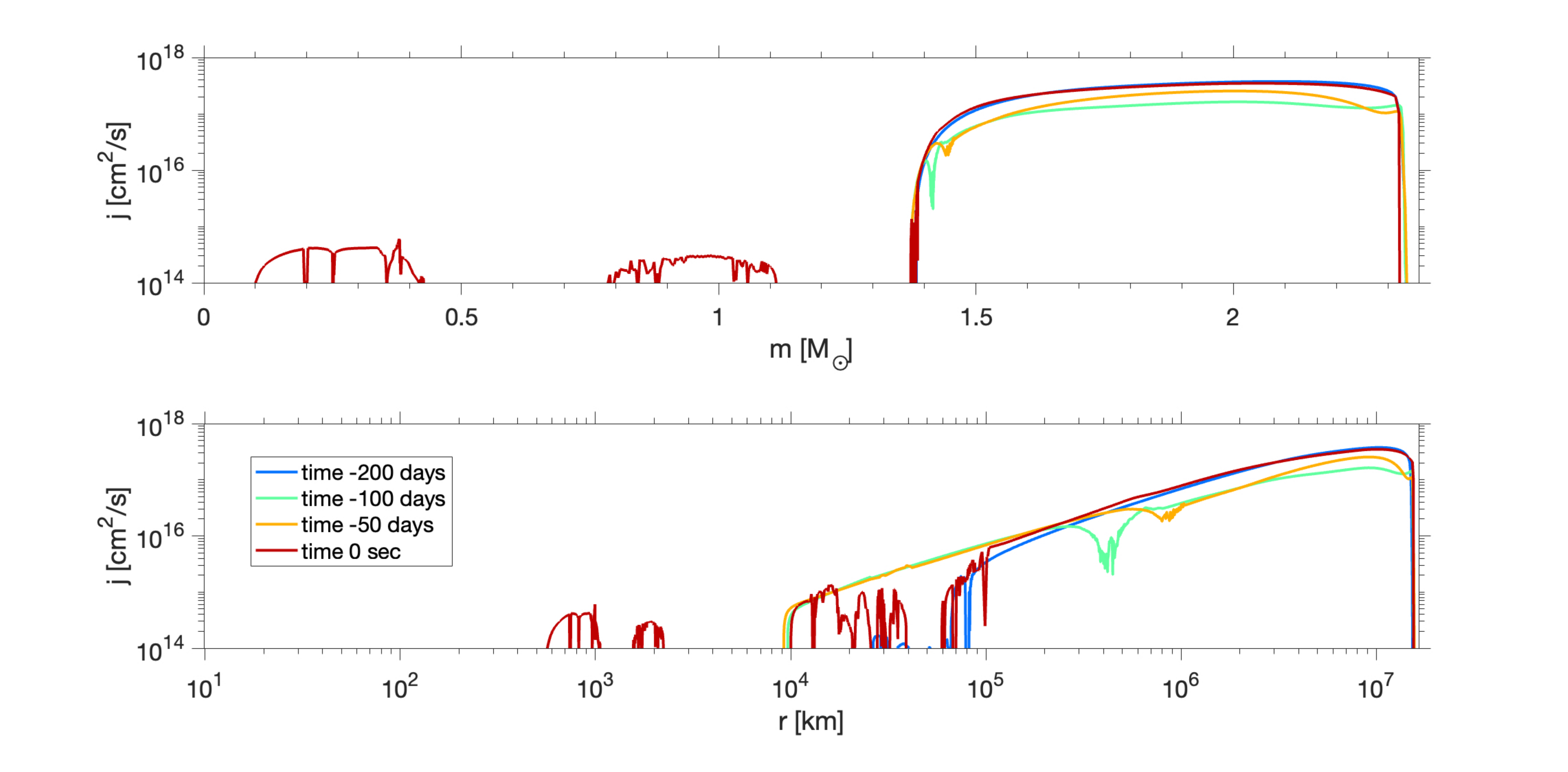} 
\caption{The specific angular momentum parameter $j$ of the M88 model in the inner zone up to the hydrogen burning layer, and at four times, as a function of mass (upper panel) and of radius (lower panel). The time $t=0$ corresponds to the last time of our simulation. Inset gives the time before the termination of the simulation. The layers that will eventually launch the exploding jets when accreted onto the newly born NS are at $m > 1.373 M_\odot$, where convection properties do not change much towards collapse.  }
\label{fig:ConvTimeM88}
\end{center}
\end{figure*}

Our goal is to explore the convective motion in the layer that we expect to explode the star according to the JJEM (section \ref{sec:Intro}). In Figure \ref{fig:ConvTimeM88} we present the specific angular momentum parameter $j$ (equation \ref{eq:j}) in the inner parts of the star and at four times; $j(m)$ in the upper panel and $j(r)$ in the lower panel. The time $t=0$ corresponds to the last time of our simulation. The convection zones in the inner core $m<1.379 M_\odot$ change substantially with time. However, the layer that we expect to explode the star as it is accreted onto the newly born NS, is where $j \ga 2.5 \times 10^{15} \cm^2 \s^{-1}$ (e.g., \citealt{ShishkinSoker2021}). The relevant layer in the M88 model is around $m=1.38 M_\odot$. In that region, the convection is robust and does not change much with time towards the end of our simulation, as evident by the overlapping lines at those mass coordinates (upper panel of Figure ~\ref{fig:ConvTimeM88}). We note the presence of convection in inner regions ($m<1M_\odot$) at the latest simulated time. However, the values of $j$ in this inner zone are too small to set an explosion according to our criterion. 

\begin{figure*}[t]
\begin{center}
\includegraphics[trim=0.5cm 0cm 0.5cm 0.57cm,width=\textwidth,scale=.6]{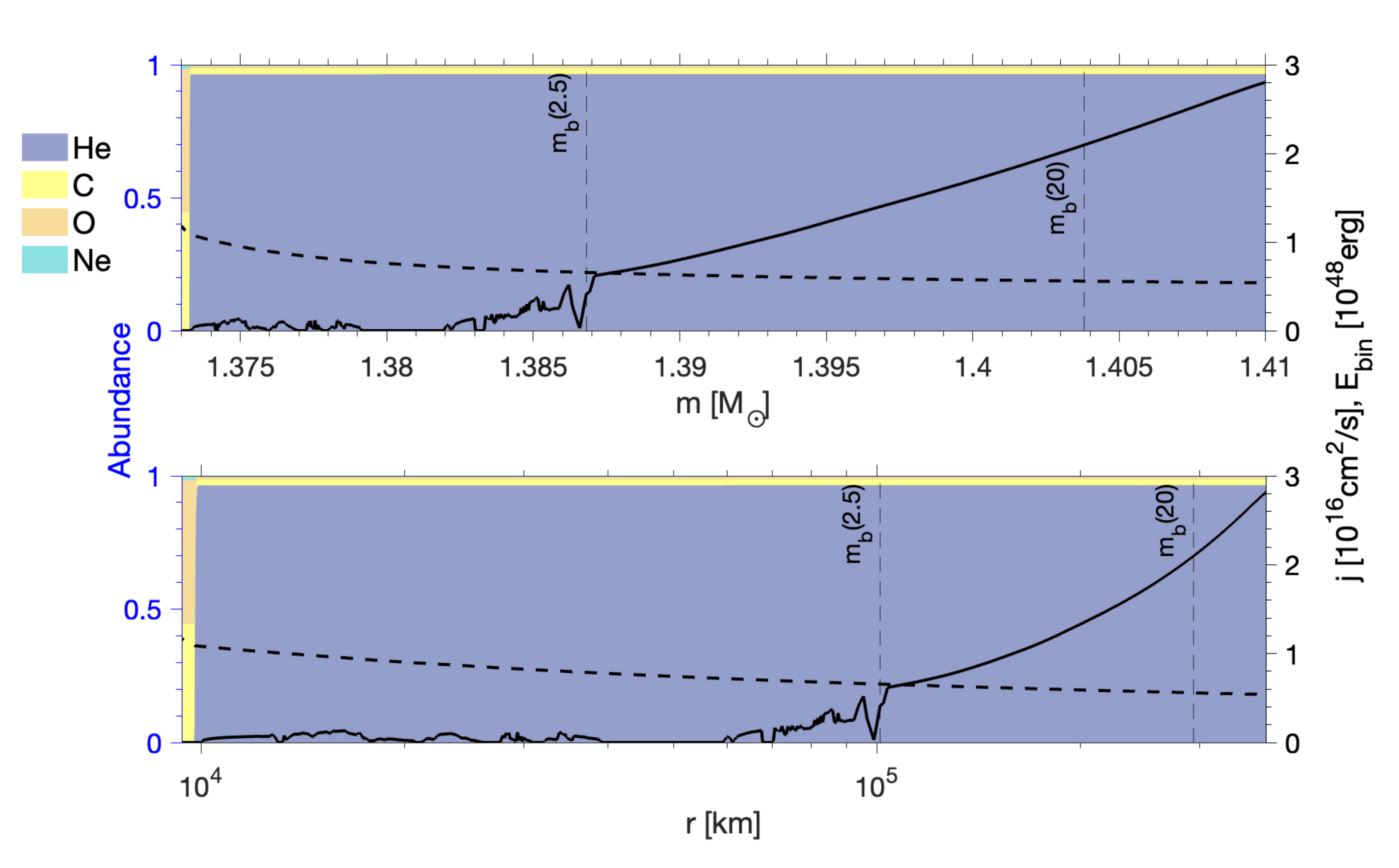} 
\caption{Zooming in on the exploding layers of model M88, $m=1.373 M_\odot - 1.410 M_\odot$. The colors show the composition (legend on left), the black-solid line is the specific angular momentum parameter $j$ (scale on right) and the black-dashed line depicts the binding energy $E_{\rm bin}$ (scale on right). 
The vertical dashed lines show the mass coordinates at which sufficient specific angular momentum parameter exists for the JJEM to operate by one of two different criteria, either $j_{\rm jje}=2.5 \times 10^{15} \cm^2 \s^{-1}$ or $j_{\rm jje}=20 \times 10^{15} \cm^2 \s^{-1}$. The exploding layer is expected to be somewhere between the two vertical lines and to include $\approx 0.001 M_\odot$.}
\label{fig:ComBinJZoomM88}
\end{center}
\end{figure*}

Simulations of CCSNe (e.g., \citealt{Jankaetal2007}) show that the inner region of $m \la 1.1 M_\odot$ collapses before the unstable region behind the stalled shock at $r \simeq 100-150 \km$ is fully developed.  We therefore expect that the relevant convective layers are those around $m \simeq 1.4 M_\odot$ which corresponds to $r \approx 10^5 \km$ in our model M88. In Figure \ref{fig:ComBinJZoomM88} we zoom on the mass zone $m=1.373 - 1.410 M_\odot$, showing composition, $j$, and the binding energy $E_{\rm bin}$. This zone is basically in the helium-rich layer. The binding energy is the binding energy of the mass above the mass or radius coordinate. Namely, it is the minimum energy that is needed to explode the stellar mass above the mass coordinate.  

According to the JJEM for FeCCSNe a value of $j \ga 2 \times 10^{15} \cm^2 \s^{-1}$ should be sufficient to form intermittent accretion disks that launch the jets (e.g., \citealt{ShishkinSoker2021}). The reason is that instabilities, mainly the spiral SASI, amplify these angular momentum fluctuations (see section \ref{sec:Intro}). \cite{ShishkinSoker2022} who study FeCCSNe consider two lower limits to the specific angular momentum in the pre-explosion convective layer to trigger an explosion,  $j_{\rm jje}=2.5 \times 10^{15} \cm^2 \s^{-1}$ and $j_{\rm jje}=5 \times 10^{15} \cm^2 \s^{-1}$. However, because of the longer free fall time and the possibility that in ECSNe the spiral-SASI and other instabilities are not as developed as in FeCCSNe, we take also a much larger value for the critical value of $j$ and consider also a limit of $j_{\rm jje}=2 \times 10^{16} \cm^2 \s^{-1}$. The latter value is comparable to the specific angular momentum required to form an accretion disk around an NS.
We mark in Figure \ref{fig:ComBinJZoomM88} by dashed vertical lines the locations in the star where the value of $j$ is one of the two criteria $j_{\rm jje}=2.5 \times 10^{15} \cm^2 \s^{-1}$ or  $j_{\rm jje}=20 \times 10^{15} \cm^2 \s^{-1}$.

We can crudely estimate the required accreted mass as follows (see earlier studies of the JJEM listed in section \ref{sec:Intro}). The binding energy at $m \simeq 1.4 M_\odot$ is $E_{\rm bin} (1.38)\simeq 7\times 10^{47} \erg$ (more on that in section \ref{sec:Models}). We assume that to explode the star the energy that the jets carry is several times the binding energy and up to tens times the binding energy, $E_{\rm jets} = f_{\rm exp} E_{\rm bin}$ with $f_{\rm exp} \approx 10$. We take the jets to carry a fraction $\eta_{\rm j} \simeq 0.1$ of the accreted mass and have a velocity of $v_{\rm j} \simeq 10^5 \km \s^{-1}$ and up to $v_{\rm j} \simeq 0.5 c$. This gives for the required accreted mass during the jet-launching episodes   
\begin{equation}
\begin{split}
M_{\rm acc} & = 0.001 
\left( \frac{f_{\rm exp}}{10} \right)
\left( \frac{E_{\rm bin}}{10^{48} \erg} \right)  \\
\times & 
\left( \frac{v_{\rm j}}{10^5 \km \s^{-1}} \right)^{-2}
\left( \frac{\eta_j}{0.1} \right)^{-1} M_\odot. 
\label{eq:Macc}
\end{split}
\end{equation}
Lower efficiency $\eta_{\rm j}$ will increase the required mass in the accreted layer during the jet-launching episodes. 

The layers we expect to explode model M88 fall inside the helium-rich shell, as the two vertical dashed lines mark on Figure \ref{fig:ComBinJZoomM88}. For the critical value of $ j_{\rm jje}=2.5 \times 10^{15} \cm^2 \s^{-1}$ 
The exploding shell is at $m_b(2.5)=1.387M_\odot$ and $r_b(2.5)=1.01 \times 10^5 \km = 0.145 R_\odot$, and for 
$j_{\rm jje}=20 \times 10^{15} \cm^2 \s^{-1}$ the layer is at $m_b(20)=1.404 M_\odot$ and $r_b(20)=2.95 \times 10^5 \km = 0.424 R_\odot$.
The binding energy of the mass above these shells are 
$E_{{\rm B},b}(2.5)=  6.57 \times 10^{47} \erg$  and 
$E_{{\rm B},b}(20)=  5.58 \times 10^{47} \erg$, respectively.  

The free fall times of these layers to the center are
$t_{{\rm ff},b}(2.5)=83 \s$ and 
$t_{{\rm ff},b}(20)=413 \s$, respectively. 
These timescales imply that the explosion of the core starts at tens of seconds to several minutes after the formation of the proto-NS (after the bounce of the collapse). These typical timescales are about two to three orders of magnitude longer than the explosion time of FeCCSNe, where the exploding layer is either near the silicon burning zone or near the oxygen burning zone at a radius of $\approx 3000 \km$ (\citealt{ShishkinSoker2021}). 
In the M88 ECSNe model that we study here the exploding layer is in the helium-burning zone. Because the shock takes hours and more to break out from the red supergiant, this extra time of several minutes does not change much for the explosion of the star. 

We conclude that the ECSN progenitor model M88 has sufficiently high values of the angular momentum parameter $j$ to explode the star, with the exploding layer at $m_{\rm b} \simeq 1.39 - 1.41 M_\odot$. This is the baryonic mass. The final NS mass will be about 15\% lower (e.g., \citealt{LattimerPrakash2001}; more in section \ref{sec:Models}). Our model did not reach the time of core collapse itself. Nonetheless, we showed that towards the collapse there is an extended convective zone that can provide the necessary angular momentum fluctuations for the JJEM to operate. This convection is likely to remain stable until collapse (Sec.~\ref{subsec:GeneralProperties88}). Even if near the core boundary there is no such a convective zone, there is such a convective zone further out. Since the density above the core is low, even if the exploding layer is at large radii as $\simeq  R_\odot$, the final NS mass is still $M_{\rm NS} \simeq 1.2-1.4 M_\odot$. We will see this also in section \ref{sec:Models} for other stellar models.  

A note on the threshold CO core mass for electron capture supernovae is in place here. While some papers on the ECSN (e.g. \citealt{Limongi_etal_2024}) cite a value of $1.415 M_\odot$, others claim to get an explosion with lower core masses, e.g., $1.34 M_\odot$ \citep{Stockingeretal2020}. The value is also highly dependent on the mixing properties (e.g. \citealt{Zha_etal_2019} who also state a CO core value of $1.36 M_\odot$ leading to an explosion). Our M88 model has a CO core mass of $M^{\rm CO}_{\rm core} = 1.375$, which is inside the range of the values mentioned above.

\section{The $M_{\rm ZAMS} = 9.1M_\odot$ model (M91)}
\label{sec:M91}

\begin{figure*}[t]
\begin{center}
\includegraphics[trim=0.5cm 0cm 0.5cm 0.57cm,width=\textwidth,scale=.66]{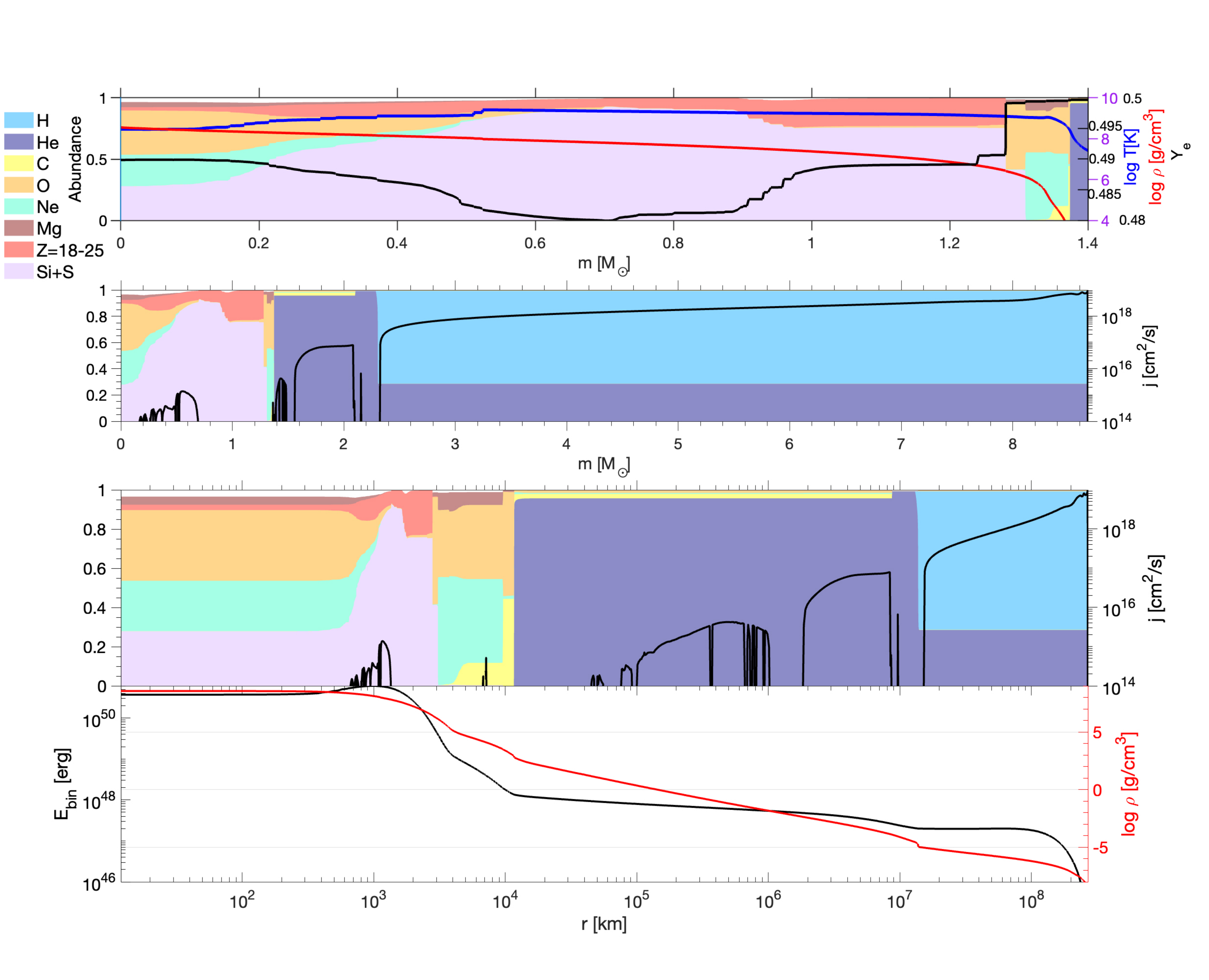} 
\caption{Similar to Figure \ref{fig:FinalCompM88} but for model M91 of $M_{\rm ZAMS} = 9.1 M_\odot$. The ranges of axes are different. 
}
\label{fig:FinalCompM91}
\end{center}
\end{figure*}

\begin{figure*}[t]
\begin{center}
\includegraphics[trim=0.5cm 0cm 0.5cm 0.57cm,width=\textwidth,scale=.6]{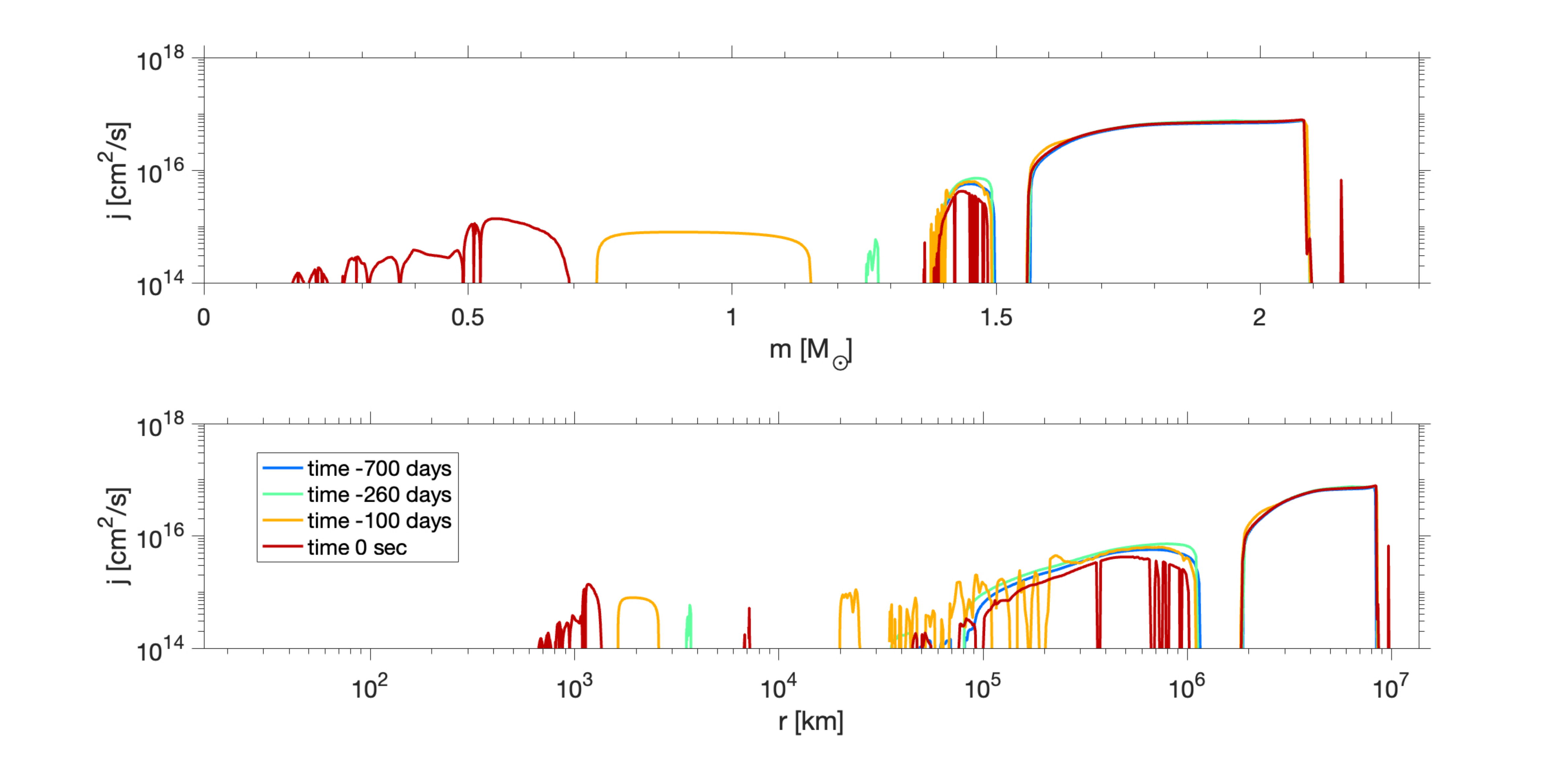} 
\caption{Similar to Figure \ref{fig:ConvTimeM88} but for model M91 of $M_{\rm ZAMS} = 9.1 M_\odot$. The ranges of axes and times of the profiles are different. 
}
\label{fig:ConvTimeM91}
\end{center}
\end{figure*}

\begin{figure*}[t]
\begin{center}
\includegraphics[trim=0.5cm 0cm 0.5cm 0.57cm,width=\textwidth,scale=.6]{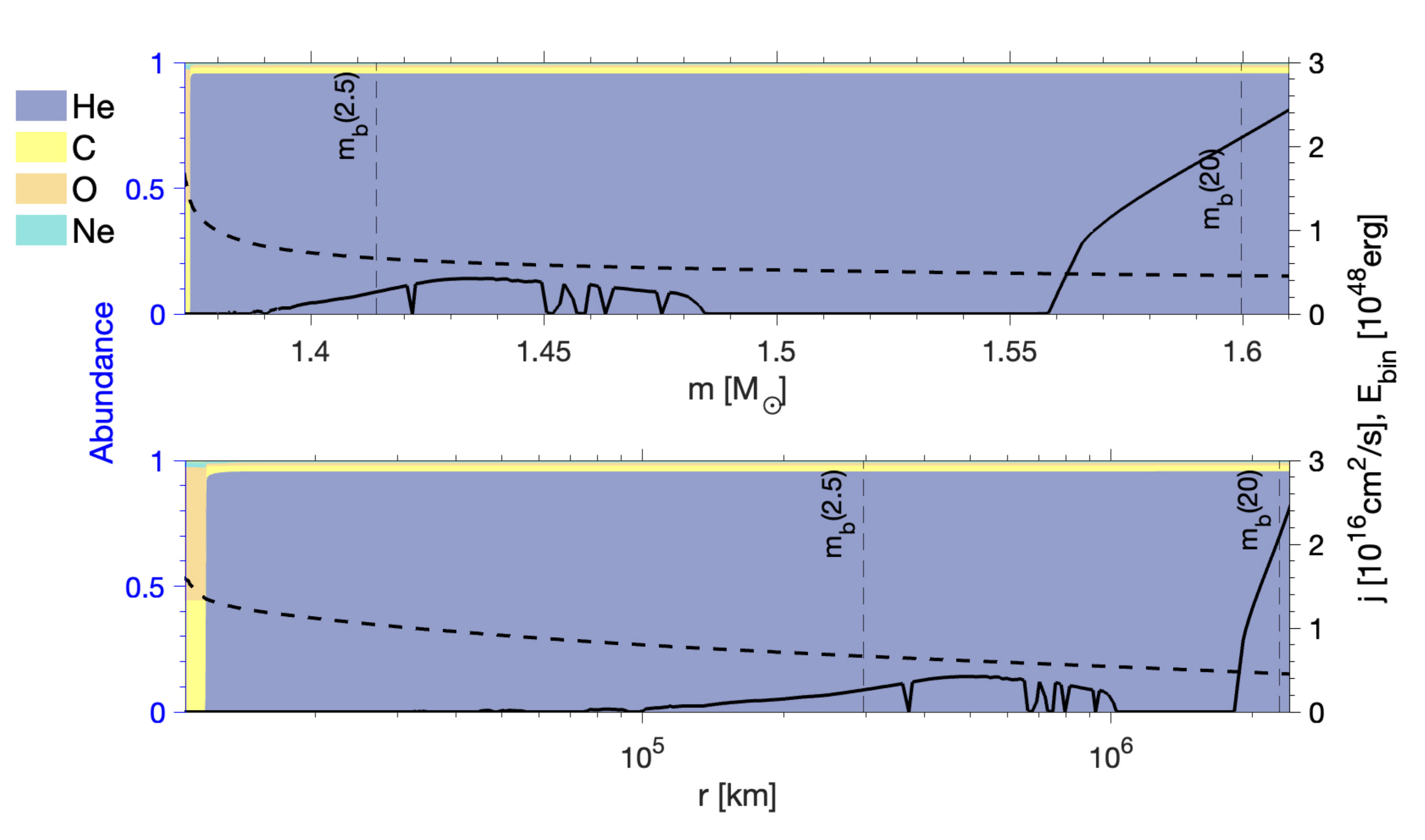} 
\caption{Similar to Figure \ref{fig:ComBinJZoomM88} but for model M91 of $M_{\rm ZAMS} = 9.1 M_\odot$. The ranges of axes are different. 
}
\label{fig:ComBinJZoomM91}
\end{center}
\end{figure*}

We simulated five models that we summarize in section \ref{sec:Models}. Here we present the details of a model with $M_{\rm ZAMS} = 9.1M_\odot$ and $z=0.0086$ which we name M91 because it differs the most from model M88, which we presented in section \ref{sec:M88}. Model M91 has a much longer free-fall time from the convective layer with the condition $j_{\rm jje}=20 \times 10^{15}$. 
In Figure \ref{fig:FinalCompM91} we present its composition and the other properties that we presented in Figure \ref{fig:FinalCompM88} for model M88, at the time when the simulation halted because of too small time steps. We present the evolution with time of the convective layers in Figure \ref{fig:ConvTimeM91}, as we did in Figure \ref{fig:ConvTimeM88} for model 88, but for different times. In Figure \ref{fig:ComBinJZoomM91} we zoom on the convective layers that we expect to explode the star, as we did for model M88 in Figure \ref{fig:ComBinJZoomM88}. 

Comparing Figure \ref{fig:ComBinJZoomM88} with Figure \ref{fig:ComBinJZoomM91}, we find that qualitatively the composition is similar in models M88 and M91. There are some small differences, like that model M91 does not have yet iron group elements in its core. Comparing the upper panels of Figures \ref{fig:ConvTimeM88} and \ref{fig:ConvTimeM91} with each other, we notice more differences. Relevant to us is that in both model M88 and model M91 there is a robust and continuous convective zone above $m \simeq 1.38M_\odot$ with high values of $j$. The convective layer $1.4 M_\odot \lesssim m \lesssim 1.5 M_\odot$ has lower values of $j$ than in M88 and it is still somewhat varying with time (Figure \ref{fig:ConvTimeM91}). The location of the accreted gas that launches the exploding jets of model M91 according to the JJEM is at $m_b(2.5)=1.414M_\odot$ for $j_{\rm jje}=2.5 \times 10^{15} \cm^2 \s^{-1}$ and $m_b(20)=1.598 M_\odot$ for $j_{\rm jje}=20 \times 10^{15} \cm^2 \s^{-1}$ (upper panel of Figure \ref{fig:ComBinJZoomM91}). We take these masses to be the baryonic mass of the NS remnant (hence the subscript `b'; more on these and the gravitational NS masses in section \ref{sec:Models}). The locations of these two zones is at $r_b(2.5) = 0.424 R_\odot = 2.95 \times 10^{5} \km$ and $r_b(20) = 3.25 R_\odot = 2.26 \times 10^{6} \km$, respectively (lower panel of Figure \ref{fig:ComBinJZoomM91}). 

From Figure \ref{fig:ComBinJZoomM91} we learn two interesting things. The first is that the accreted material that launches the jets originates in a helium layer, as studied by \cite{GilkisSoker2016} for FeCCSNe, and not from the hydrogen-rich envelope as studied by, e.g., \cite{AntoniQuataert2022} and \cite{AntoniQuataert2023}, for FeCCSNe. Namely, there is no `failed supernova' here that is exploded only by the outer envelope layers. In any case, according to the JJEM there are no `failed supernovae' (because jets explode all core-collapsing massive stars), a claim supported by recent observations \citep{ByrneFraser2022, StrotjohannOfekGalYam2023} and theoretical considerations (see review by \citealt{Soker2024Rev}). 

The second interesting result is that for $r_b(20) = 3.25 R_\odot$ the free fall time from that radius to the center with $m_b(20)=1.598 M_\odot$ is $t_{\rm ff,b}(20) =8217 \s = 2.3~\rm h$. This is a much longer time than the explosion of FeCCSNe.  We first note that we did not reach core collapse. We expect that at the time of the actual core collapse, the radius of the accreted zones that power the exploding jets will be smaller. We discuss this point in section \ref{sec:Models}. Even if the exploding jets are launched a few hours after the core collapses, the explosion energy and observed event might be similar to that of a much shorter explosion. The reason is that for the expected explosion energy in these cases, $E_{\rm ex} \simeq 10^{49} - 10^{50} \erg$, and the extended hydrogen-rich envelope, the stellar radius is two orders of magnitude larger than the radius that supplies the gas to launch the jets and the shock wave transverses the star in about a day, longer than the free-fall time. Future hydrodynamical simulations should reveal the properties of such exploding stars.   

We end this section by noting that some recent results suggest that even FeCCSNe in the frame of the delayed neutrino explosion mechanism require several seconds of explosion process (e.g., \citealt{Bollig_etal_2021, BurrowsVartanyan_2021_Nature}), rather than only one second, to reach observed explosion energies. Overall, the CCSN explosion process, in some cases, might be much longer than the canonical one-second explosion timescale (which is the dynamical timescale of the collapsing inner core). This, in turn, allows the accretion of core layers outside the silicon core. In any case, the final mass of the proto-NS in ECSN or FeCCSN simulations can extend the $\simeq 1.2-2 \Mo$ range, implying the accretion of outer core layers in most models.

\section{Other models}
\label{sec:Models}

We simulated three more models that are likely to explode as ECSNe and examined the relevant parameters to the JJEM.  We list the initial masses and metalicity of the five models in the second and third rows of Table \ref{Tab:Table1}. The other rows list properties at the last time of the simulation of each model. We recall that the last time of each simulation was when the numerical time steps became too small to continue the simulation. We did not reach the core collapse phase in any of the models. 

\begin{table*}
\scriptsize
\caption{Some properties of ECSN progenitors}
\begin{tabular}{|l|c|c|c|c|c|c|c|}
\hline
 Quantity & Units & M88 & M89 & M91 & M93 & M945\\
\hline
$M_{\rm ZAMS}$ & $M_\odot$ & 8.8 & 8.9 & 9.1 & 9.3 & 9.45 \\
\hline
$z_{\rm ZAMS}$ &   & 0.004 & 0.0057 & 0.0086 & 0.012 & 0.014 \\
\hline
$m_{\rm b}(2.5)$ & $M_\odot$ & 1.387 & 1.376 & 1.414 & 1.381 & 1.413 \\
$M_{\rm NS}$ & $M_\odot$ & 1.2598 & 1.2513 & 1.2822 & 1.2546 & 1.2814 \\
\hline
$m_{\rm b}(20)$ & $M_\odot$ & 1.404 & 1.577 & 1.598 & 1.417 & 1.523 \\
$M_{\rm NS}$ & $M_\odot$ & 1.2739 & 1.4151 & 1.4316 & 1.2845 & 1.3712 \\
\hline
$r_{\rm b}(2.5)$ & $R_\odot$ & 0.145 & 0.037 & 0.424 & 0.093 & 0.472 \\
\hline
$r_{\rm b}(20)$ & $R_\odot$ & 0.424 & 3.275 & 3.254 & 0.628 & 2.669 \\
\hline
$t_{{\rm ff,b}}(2.5)$ & \s & 83 & 11 & 411 & 43 & 483 \\
\hline
$t_{{\rm ff,b}}(20)$ & \s & 413 & 8350 & 8217 & 740 & 6250 \\
\hline
$E_{{\rm B,b}}(2.5)$ & $10^{47} \rm erg$ & 6.57 & 9.71 & 6.66 & 6.15 & 4.46 \\
\hline
$E_{{\rm B,b}}(20)$ & $10^{47} \rm erg$ & 5.58 & 4.35 & 4.6 & 4.04 & 3.03 \\
\hline
$^{16}$O &   & 0.0911 & 0.15 & 0.112 & 0.2705 & 0.178 \\
$^{20}$Ne &   & 0.061 & 0.0561 & 0.0665 & 0.1939 & 0.1162 \\
$^{24}$Mg &   & 0.0081 & 0.01 & 0.01 & 0.0259 & 0.0171 \\
$^{28}$Si$+^{32,34}$S &   & 0.5245 & 0.5861 & 0.6004 & 0.3715 & 0.538 \\
$^{56}$Fe &   & 0.06503 & $3.9\times 10^{-4}$ & $5.9\times 10^{-4}$ & $3.6\times 10^{-6}$ & $8.0\times 10^{-7}$ \\
\hline
$\rm Y_{e,min}$ &   & 0.481391 & 0.484083 & 0.479937 & 0.490685 & 0.484523 \\
\hline
$m(\rm Y_{e,min})$ & $M_\odot$ & 0.649 & 0.893 & 0.701 & 0.945 & 0.616 \\
\hline
$\log\rho_{c}$ & $\g \cm^{-3}$ & 8.505 & 8.462 & 8.556 & 8.527 & 8.316 \\
\hline
\end{tabular}
\label{Tab:Table1}

\begin{flushleft}
\small 
Notes: Properties of the five simulated models, from top to bottom below the name of each model: $M_{\rm ZAMS}$: the initial stellar mass on the ZAMS; $z_{\rm ZAMS}$: initial metalicity. Results from stellar evolution simulations at the last time of the simulation in each model are as follows. $m_{\rm b}(j_{\rm jje})$: the radius inner to which there is a layer of $0.001 M_\odot$ with specific angular momentum above the critical value $j_{\rm jje}=2.5 \times 10^{15} \cm^2 \s^{-1}$ or $j_{\rm jje}=20 \times 10^{15} \cm^2 \s^{-1}$. $M_{\rm NS}$: The final gravitational mass of the NS according to equation (\ref{eq:Mgrav}).  $r_{\rm b}(j_{\rm jje})$: The radius at mass coordinate $m_{\rm b}(j_{\rm jje})$; $t_{\rm ff}(j_{\rm jje})$: The free fall time from $r_{\rm b}$ to the center; $E_{{\rm B,b}}$: the binding energy of the stellar mass above $m_{\rm b}$.
Five rows give the fraction of the respective isotopes in the ONeMg core. The final three rows list the minim electron fraction in the core $Y_{\rm e, min}$, the corresponding mass coordinate $m(Y_{\rm e, min})$, and the logarithm of the density in the center $\log \rho_c$.
\end{flushleft}

\end{table*}

To estimate the location of the convective layers in the star that supply the material to the stochastic accretion disk that launches the exploding jittering jets, we use two criteria.

The first criterion is that the convective specific angular momentum parameter (equation \ref{eq:j}) should be larger than a critical value $j_{\rm jje}$. We note that the exploding layer should be at a mass coordinate $m>1.1 M_\odot$. The reason is that the spiral-SASI should have time to develop above the newly born NS for it to amplify the convective seed angular momentum perturbations.
As we described in section \ref{subsec:ConvectiveZones88}
we examine two values of $j_{\rm jje}=2.5 \times 10^{15} \cm^2 \s^{-1}$ and $j_{\rm jje}=20 \times 10^{15} \cm^2 \s^{-1}$. We expect the exploding layer to be somewhere in between these two layers. 

The second criterion is that the mass in the exploding layer should be $\simeq 0.001 M_\odot - 0.01M_\odot$ so that the mass in the two jets with a terminal velocity of $v_{\rm j} \simeq 10^5 \km \s^{-1}$ carry the required explosion energy (equation \ref{eq:Macc}). 

We take the mass coordinate $m_{\rm b}(j_{\rm jje})$ inner to which there is a mass layer of $0.001 M_\odot$ with $j>j_{\rm jje}$ to be the baryonic mass of the neutron star $m_{\rm b}$. The final gravitational mass of the NS, $M_{\rm NS}$, is lower by about $\simeq 10-15\%$, and is given by (from \citealt{LattimerPrakash2001} as used by \citealt{Sukhbold_etal_2016})
\begin{equation}
{\scriptstyle
M_{\rm NS}=\frac{5}{\tilde{G}}\left(\sqrt{\left(1+0.5\tilde{G}M_b \right)^2 +0.4\tilde{G}M_b}-1-0.5\tilde{G}M_b\right), 
}
\label{eq:Mgrav}
\end{equation}
where $\tilde{G} \equiv GM_\odot /(R_{\rm NS}~c^2)=0.123(R_{\rm NS}/12 \km)^{-1}$ and $M_{b}~[M_\odot]$ is the baryonic mass at collapse.

We record the pre-collapse radius of the mass coordinate $m_{\rm b}(2.5)$ for $j_{\rm jje}=2.5 \times 10^{15} \cm^2 \s^{-1}$ and $m_{\rm b}(20)$ for $j_{\rm jje}=20 \times 10^{15} \cm^2 \s^{-1}$. We expect the layer that feeds the exploding jets to be somewhere in between these two values. We present the baryonic and final NS masses for the two values of $j_{\rm jje}$ in the fourth to seventh rows of Table \ref{Tab:Table1}. 
In the following rows, we list the radii of the above two layers, and then the free fall time from these two layers to the center, $t_{{\rm ff,b}}$. We then list the binding energy of the mass above that radius $E_{{\rm B,b}}$ for the two values of $j_{\rm jje}$. In five rows we list the fractional composition of some isotopes in the ONeMg core, i.e., inner to the point where $M \simeq 1.35 M_\odot$.
The final three rows list the minimum electron fraction in the core $Y_{\rm e, min}$, the mass coordinate of this minimum $m(Y_{\rm e, min})$, and the logarithm of the density in the center. 

There are two prominent outcomes from Table \ref{Tab:Table1} that are relevant to our study in the frame of the JJEM. The first is that the final NS mass of an ECSN in all cases according to the criteria we use falls in the range of $1.25 M_\odot < M_{\rm NS} < 1.43 M_\odot$. The very low density above the pre-collapse core ensures that even if the exploding layer is further out in the core the final NS mass will be $\simeq 1.4 M_\odot$, compatible with observations. We note that \cite{Stockingeretal2020} and \cite{Radice_etal_2017} who simulated neutrino-driven explosion found the final NS mass in their ECSN models to be in the range of $1.195-1.23 M_\odot$ and $1.188 M_\odot \lesssim M_{\rm NS} < 1.4 M_\odot $, respectively. The range we find here for the final NS masses might change following more accurate simulations and the inclusion of pre-collapse core rotation. However, we expect that the final NS masses that the JJEM predicts will stay somewhat larger than what the neutrino-driven explosion mechanism predicts. 

The second outcome is that the explosion by jets might occur with a substantial time delay after the core collapse starts. In FeCCSNe the explosion occurs within a second to several seconds after the core collapses. In our simulations, we find that the free fall time of the exploding layer can be from about a minute to a few hours, implying such a delay from the core collapse to the explosion. We recall that we did not reach the time of core collapse due to numerical difficulties. The exploding layer may be somewhat closer at the time of core collapse, but we expect the delay to be longer than in FeCCSNe. The implication on nucleosynthesis and on other explosion properties is a subject of future simulations. We do not expect the observed lightcurve to differ much from a case with a short delay of only several seconds because the delay of the free fall time, even if a few hours, is much shorter than the time the shock crosses the entire star. 

We conclude that the pre-collapse cores of ECSN progenitors have convective layers that can feed the intermittent accretion disks that launch jittering jets according to the JJEM. 

\section{Summary}
\label{sec:Summary}

Studies in recent years show that many, and probably most if not all, CCSNe are exploded by jets (e.g., \citealt{Soker2024CFpoint} and \citealt{Soker2024Rev} for a very recent review and earlier references therein). The jet-driven explosion mechanism that can work in most CCSNe is the JJEM. Earlier studies of the JJEM concentrate on FeCCSNe (section \ref{sec:Intro}). Therefore, we aim to examine whether the progenitors of ECSNe also have the convective layers that can supply the stochastic angular momentum to the intermittent accretion disks around the newly born NS that launch the exploding jets. To that goal, we simulated five stellar models summarized in Table \ref{Tab:Table1}.

For a pre-collapse convective layer, the `exploding layer', to supply the stochastic angular momentum and mass of the intermittent accretion disks in CCSNe, we demand a minimum specific angular momentum parameter $j>j_{\rm jje}$ (equation \ref{eq:j}), and minimum mass $M_{\rm acc}$ according to equation (\ref{eq:Macc}). We examined two values of $j_{\rm jje} = 2.5 \times 10^{15} \cm^2 \s^{-1}$ and $j_{\rm jje} = 20 \times 10^{15} \cm^2 \s^{-1}$; we expect the exploding layer to be somewhere in between these two. 
This threshold value is a conjecture of the JJEM used by studies of the JJEM in FeCCSNe (see section \ref{sec:Two Mechanism}), which we adopted here for ECSNe. It has yet to be validated by numerical simulations. We hope the new findings of point-symmetric CCSN remnants will motivate such studies.

We found that the exploding layer resides in the helium-rich shell above the CO core (Figures \ref{fig:ComBinJZoomM88} and \ref{fig:ComBinJZoomM91}). This is different than in FeCCSNe where the exploding layers are near the silicon-burning shell or the oxygen-burning shell of the pre-collapse core (e.g., \citealt{ShishkinSoker2021}). We found that the progenitors of ECSNe have convective layers in their helium-rich zone that fulfill the demands of the JJEM to set an explosion when the newly born NS accretes them. 

We find that the exploding layer resides at a distance of $r_{\rm b} \approx 0.05-3 R_\odot$ (Figures \ref{fig:ComBinJZoomM88} and \ref{fig:ComBinJZoomM91}, and Table \ref{Tab:Table1}). This implies a free-fall time to the center of $f_{\rm ff,b} \approx {\rm few \times~10~sec} - {\rm few\times~hour}$. Because our simulations stop for numerical reasons before the core collapse phase, the actual radius of the exploding layers at the core collapse might be smaller. Nonetheless, the explosion phase might occur minutes to hours after core collapse, but still much shorter than the time the shock wave crosses the giant star. Therefore, the CCSN lightcurve is expected to be similar to that where the delay is only a few seconds. Other effects should be the topic of future simulations. 

Overall, our results that ECSN progenitors have convective layers that can serve as the exploding layers of the JJEM even at the lower mass range of CCSNe (namely, ECSNe) add to the growing support \citep{Soker2024Rev} of the JJEM as the main explosion mechanism of massive stars.

\section*{Acknowledgments}

We thank an anonymous referee for detailed and helpful comments and suggestions. This research was supported by a grant from the Israel Science Foundation (769/20).

\appendix

\makeatletter
\renewcommand\@makefnmark{\hbox{\@textsuperscript{\normalfont\color{white}\@thefnmark}}}
\renewcommand\@makefntext[1]{%
  \parindent .5em\noindent
            \hb@xt@1.8em{%
                \hss\normalfont\@thefnmark}#1}
\makeatother

\begin{figure*}[b]
\begin{center}
\includegraphics[trim=0.5cm 2.5cm 0.5cm 10cm,width=0.74\textwidth]{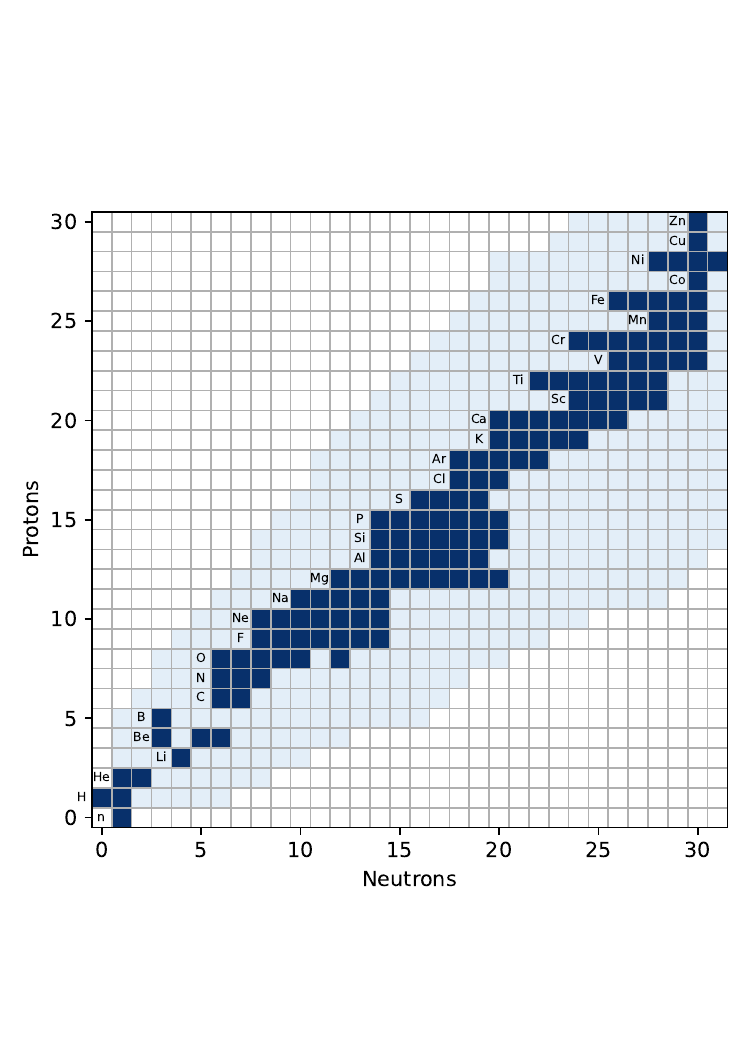} 
\caption{The nuclear network used for the simulations of the ECSN progenitors. Horizontal (vertical) axis is the nuclii neutron (proton) number. In dark blue are the isotopes we included in the \mesa~network, with light blue being an exhaustive set of known nuclides. Note the presence numerous magnesium, neon, fluorine and oxygen isotopes that play a role in the different electron capture chains.
Image compiled and data extracted using the $\rm mesa\_net\_reader$ tool on Github\textsuperscript{\ref{footnoteAppendix}}.}
\label{fig:nucNet}
\end{center}
\end{figure*}

\color{black}{\footnote[1]{\label{footnoteAppendix}\href{https://github.com/SDcodenum/mesa_net_reader}{mesa\_net\_reader} is a tool that reads mesa network files and extracts a list of all the isotopes included, for easy plotting. Also includes an example plotting scheme. Available on GitHub at \href{https://github.com/SDcodenum/mesa_net_reader}{https://github.com/SDcodenum/mesa\_net\_reader}}}

\end{document}